\documentstyle[aps,prb,preprint]{revtex}
\begin{document}
\draft
\title{p-wave and d-wave superconductivity in quasi-2D metals}
\author{P. Monthoux and G.G. Lonzarich}
\address{Cavendish Laboratory, University of Cambridge
\\Madingley Road, Cambridge CB3 0HE}
\date{\today}
\maketitle

\begin{abstract}

We compare predictions of the mean-field theory of superconductivity
for nearly antiferromagnetic and nearly ferromagnetic metals in two
dimensions. The calculations are based on a parametrization of the
effective interaction arising from the exchange of magnetic 
fluctuations. The Eliashberg equations for the transition temperature
are solved including the full momentum dependence of the electron 
self-energy. The results show that for comparable parameters d-wave
singlet pairing in nearly antiferromagnetic metals is generally much 
stronger than p-wave triplet pairing in nearly ferromagnetic metals in 
quasi two dimensions. The relevance to the layered materials, and in 
particular $Sr_2RuO_4$ that exhibits p-wave triplet pairing, is discussed.

\end{abstract}


\narrowtext

\section{Introduction}

There is growing experimental evidence of anisotropic forms of 
superconductivity in the quasi two-dimensional perovskite oxides. 
Energy gaps of d-wave character have been established for some of the 
copper oxides that have strongly enhanced antiferromagnetic 
susceptibilities and high superconducting transition temperatures (of
the order of $100^\circ K$)\cite{Hardy,Wollman,VanHarlingen,Kirtley1,Kirtley2}. 
On the other hand, p-wave spin-triplet pairing
provides a better understanding of the experimental data in the ruthanate
$Sr_2RuO_4$ that appears to be close to ordering ferromagnetically and 
becomes superconducting only at low 
temperature (of the order of $1^\circ K$)
\cite{Agterberg,Rice,Baskaran,Maeno,Luke,Rieseman,Ishida,Mackenzie,Yoshida}.

Numerous mechanisms have been proposed for anisotropic superconductivity, 
especially in the cuprates. One of the most extensively investigated 
theoretically is based on a magnetic interaction arising via the exchange of 
enhanced antiferromagnetic spin-fluctuations
\cite{Miyake,Doug1,Doug2,Moriya,Monthoux}. Though not entirely without 
difficulties, this mechanism correctly anticipated from the beginning the
d-wave symmetry of the order parameter observed in some of the copper oxides. 
Moreover, when treated in the mean-field Eliashberg theory with full 
momentum dependence of the electron self-energy, it provided an account of 
the high transition temperatures in the cuprates, in terms of 
parameters determined independently from normal state properties alone.

In this paper we include the case where, in contrast 
to the cuprates, a magnetic interaction between electron quasiparticles 
arises from the exchange of ferromagnetic instead of antiferromagnetic 
spin-fluctuations in quasi two-dimensional (2D) compounds. Our 
calculations differ from those previously reported\cite{Levin,Fay}
for p-wave triplet pairing in the following ways: (i) they concern quasi 2D 
rather than 3D systems, (ii) employ a non-parabolic band structure which 
has potential relevance to real compounds, and (iii) make use of the 
full Green's function in place of a simple pole approximation for the 
propagator. The latter (iii) takes a better account of the momentum
dependence of the electron self-energy and was found to be important 
in the nearly antiferromagnetic 2D systems\cite{Monthoux}.
Comparisons of the mean-field Eliashberg equations for nearly 
ferromagnetic and nearly antiferromagnetic metals with a single 2D Fermi 
surface are presented for a range of parameters defining the magnetic 
interaction in potentially realistic cases. The results show that the 
incipient ferromagnets are expected to have p-wave (spin-triplet) pairing
and transition temperatures that are much lower than in the nearly 
antiferromagnetic metals for otherwise similar conditions. A physical
interpretation of the numerical analyses is given together with a discussion 
of the possible relevance of the magnetic interaction model for $Sr_2RuO_4$.
The mean-field analysis is intended as a first step toward a more complete 
treatment of superconductivity in highly correlated electron systems. It
may also serve as a possible guide to future experiments to test for 
the existence of magnetically mediated superconductivity in general.

The outline of the paper goes as follows. In the next section we describe the 
model and computational method used in this work. In section III, we describe 
the results of the numerical calculations for both ferromagnetically and 
antiferromagnetically correlated metals. Section IV contains further discussion
while our conclusions are presented in the final section.

\section{Model}

We consider quasiparticles on a two-dimensional square lattice. 
We assume that the dominant scattering mechanism is of magnetic 
origin and postulate the following low-energy effective action 
for the quasiparticles:

\begin{eqnarray}
S_{eff} & = & \sum_{{\bf p},\alpha}\int_0^\beta d\tau 
\psi^\dagger_{{\bf p},\alpha}(\tau)\Big(\partial_\tau + 
\epsilon_{\bf p} - \mu\Big) \psi_{{\bf p},\alpha}(\tau) \nonumber\\
& & - {g^2\over 6}\sum_{\bf q}\int_0^\beta 
d\tau \int_0^\beta d\tau' \chi({\bf q},\tau-\tau')
{\bf s}({\bf q},\tau)\cdot {\bf s}(-{\bf q},\tau')
\label{Seff}
\end{eqnarray}

\noindent The spin density ${\bf s}({\bf q},\tau)$ is given by

\begin{equation}
{\bf s}({\bf q},\tau) \equiv \sum_{{\bf p},\alpha,\gamma} 
\psi^\dagger_{{\bf p} + {\bf q},\alpha}(\tau){\bf \sigma}_{\alpha,\gamma}
\psi_{{\bf p},\gamma}(\tau)
\label{Spin}
\end{equation}

\noindent where $\bf \sigma$ denotes the three Pauli matrices. The 
quasiparticle dispersion relation is

\begin{equation}
\epsilon_{\bf p} = -2t(\cos(p_xa) + \cos(p_ya)) - 4t'\cos(p_xa)\cos(p_ya)
\label{eps}
\end{equation}

\noindent with hopping matrix elements t and t'. $\mu$ denotes the chemical 
potential, $\beta$ the inverse temperature, $ g^2$ the coupling constant
and $\psi^\dagger_{{\bf p},\sigma}$ and $\psi_{{\bf p},\sigma}$ are
Grassmann variables. In the following we shall measure temperatures, 
frequencies and energies in the same units. Having in mind 
a possible connection to $Sr_2RuO_4$, we shall model the sheet of
the Fermi surface of that material thought to be the most relevant for 
superconductivity\cite{Agterberg,Mazin2} by choosing t'=0.45t. With an 
average Fermi wavevector of $k_F \approx 0.7\AA^{-1}$ and a lattice constant 
$a = 3.86\AA$, Luttinger's theorem gives a doping $n \approx 1.1$. 
In the following, we shall adopt the value $n = 1.1$. The Fermi surface 
is shown in fig.(1).

Previous studies of the dependence of the critical temperature on the ratio 
t'/t and doping level\cite{Nakamura,MP2} have shown the relative insensitivity 
of $T_c$ to small changes in these parameters. Therefore, a more realistic 
description of the Fermi surface sheet of $Sr_2RuO_4$ is not expected to 
alter our conclusions. We also note that deviations from the assumed 2D 
form of the Fermi surface sheet is found experimentally to be small.

Our model assumes that the coupling parameter g is constant. The $\bf q$ 
dependence in the simplest case arises from the atomic form factor. For 
tight binding bands the latter is local in space and this leads to a weak 
dependence of g on $\bf q$. Moreover, near a magnetic instability the dominant
$\bf q$ dependence of the interaction is expected to arise from 
$\chi({\bf q},\omega)$, rather than that of g.

The retarded generalized magnetic susceptibility $\chi({\bf q},\omega)$ 
that defines the effective interaction, Eq.~(\ref{Seff}), is assumed to take 
the phenomenological form 

\begin{equation}
\chi({\bf q},\omega) = {\chi_0\kappa_0^2\over \kappa^2 + \widehat{q}^2 
- i{\omega\over \eta(\widehat{q})}}
\label{chiML}
\end{equation}

\noindent $\kappa$ and $\kappa_0$ are the inverse correlation 
lengths (in units of $a^{-1}$) with and without of strong magnetic 
correlations respectively. Let 

\begin{equation}
\widehat{q}_{\pm}^2 = 4 \pm 2(\cos(q_xa)+\cos(q_ya))
\label{qdef}
\end{equation}

In the case of ferromagnetic correlations, the parameters $\widehat{q}^2$ 
and $\eta(\widehat{q})$ are defined as

\begin{eqnarray}
\widehat{q}^2 & = & \widehat{q}_{-}^2 \\
\eta(\widehat{q}) & = & T_{sf}\widehat{q}_{-}
\label{ferro}
\end{eqnarray}

\noindent where $T_{sf}$ is a characteristic spin-fluctuation 
temperature. We shall also investigate antiferromagnetic 
correlations, in which case these parameters take the form

\begin{eqnarray}
\widehat{q}^2 & = & \widehat{q}_{+}^2 \\
\eta(\widehat{q}) & = & T_{sf}\widehat{q}_{-}
\label{antiferro}
\end{eqnarray}

The spin-fluctuation propagator on the imaginary axis, 
$\chi({\bf q},i\nu_n)$ is related to the imaginary part of the 
response function $Im\chi({\bf q},\omega)$, Eq.~(\ref{chiML}), 
via the spectral representation

\begin{equation}
\chi({\bf q},i\nu_n) = -\int_{-\infty}^{+\infty}{d\omega\over \pi}
{Im\chi({\bf q},\omega)\over i\nu_n - \omega}
\label{chi_mats}
\end{equation}

\noindent To get $\chi({\bf q},i\nu_n)$ to decay as $1/\nu_n^2$ as
$\nu_n \rightarrow \infty$, as it should, we introduce a cutoff 
$\omega_0$ and take $Im\chi({\bf q},\omega) = 0$ for $\omega 
\geq \omega_0$. A natural choice for the cutoff is $\omega_0 
= \eta(\widehat{q})\kappa_0^2$. We have checked that our results for
the critical temperature are not sensitive to the particular choice 
of $\omega_0$ used.

The two-dimensional Eliashberg equations for the critical 
temperature $T_c$ in the Matsubara representation reduce, for the
effective action Eq.~(\ref{Seff}), to

\begin{equation}
\Sigma({\bf p},i\omega_n) = g^2{T\over N}\sum_{\Omega_n}
\sum_{\bf k}\chi({\bf p}-{\bf k},i\omega_n-i\Omega_n)
G({\bf k},i\Omega_n)
\label{Sigma}
\end{equation}

\begin{equation}
G({\bf p},i\omega_n) = {1\over i\omega_n - (\epsilon_{\bf p}-\mu) 
- \Sigma({\bf p} ,i\omega_n)}
\label{Green}
\end{equation}

\begin{eqnarray}
\lambda(T)\Phi({\bf p},i\omega_n) & = & 
\Bigg[{{g^2\over 3} \atop -g^2}\Bigg]
{T\over N}\sum_{\Omega_n}\sum_{\bf k}
\chi({\bf p} - {\bf k},i\omega_n - i\Omega_n)
|G({\bf k},i\Omega_n)|^2 \Phi({\bf k},i\Omega_n) \nonumber \\
\lambda(T) & = & 1 \longrightarrow T = T_c
\label{Gap}
\end{eqnarray}

\noindent where $\Sigma({\bf p},i\omega_n)$ is the quasiparticle 
self-energy, $G({\bf p},i\omega_n)$ the one-particle Green's 
function and $\Phi({\bf p},i\omega_n)$ the anomalous self-energy.
$\epsilon_{\bf p}$ is the bare quasiparticle spectrum, 
Eq.~(\ref{eps}), $\mu$ the chemical potential that is adjusted
to give an electron density of $n = 1.1$, and $N$ the total
number of allowed wavevectors in the Brillouin Zone. In Eq.~(\ref{Gap}),
the prefactor $g^2/3$ is for triplet pairing while the
prefactor $-g^2$ is appropriate for singlet pairing. Only the
longitudinal spin-fluctuation mode contributes to the pairing 
amplitude in the triplet channel and gives rise to an attractive
interaction. Both transverse and longitudinal spin-fluctuation 
modes contribute to the pairing amplitude in the singlet channel
and give an interaction which is repulsive in reciprocal space with 
a peak at ${\bf Q} = (\pi/a,\pi/a)$. When Fourier transformed, such an 
potential is repulsive on one sublattice (even sites) and attractive 
on the other (odd sites). All three modes contribute to 
the quasiparticle self-energy.

The momentum convolutions in Eqs.~(\ref{Sigma},\ref{Gap}) are 
carried out with a Fast Fourier Transform algorithm on a 
$128\times 128$ lattice. The frequency sums in both the 
self-energy and linearized gap equations are treated with
the renormalization group technique of Pao and 
Bickers\cite{PaoBickers}. We have kept between 8 and 16 Matsubara 
frequencies at each stage of the renormalization procedure, 
starting with an initial temperature $T_0 = 0.4t$ and cutoff
$\Omega_c \approx 20t$. The renormalization group acceleration
technique restricts one to a discrete set of temperatures 
$T_0 > T_1 > T_2 \dots$. The critical temperature at which 
$\lambda(T) = 1$ in Eq.~(\ref{Gap}) is determined by linear 
interpolation. The savings in computer time and memory 
requirements afforded by this technique allowed us to study 
a wide range of temperatures and spin-fluctuation spectrum 
parameters. 

\section{Results}

The dimensionless parameters at our disposal are 
$g^2\chi_0/t$, $T_{sf}/t$, $\kappa_0$ and $\kappa$. 
It is found experimentally that $T_{sf}\kappa_0^2 \approx const$, 
and we shall use this relation to eliminate one parameter from 
the set and pick a representative value of the product 
$T_{sf}\kappa_0^2$. A value of $T_{sf} = {2\over 3} t$ 
corresponds to about $1000^\circ K$ for a bandwidth of 
1 eV while a value of $\kappa_0^2 \approx 12$ is representative
of what one obtains from a Lindhard function with 2D parabolic
bands for a Fermi momentum of about $0.7\AA^{-1}$.

The parameters of the model can in principle be inferred from the 
electronic structure, the dynamical magnetic susceptibility, and 
the resistivity in the normal state. The resistivity in particular
may be used to estimate the dimensionless coupling parameter $g^2\chi_0/t$,
the value of which is between 10 and 20 for the simplest RPA approximation
for the magnetic interaction potential.

The results of our numerical calculations of the mean-field
critical temperature $T_c$ in the case of a nearly ferromagnetic 
metal are shown in figs.(2),(3) and (4) for various values of 
the characteristic spin-fluctuation temperature $T_{sf}$. We 
find an instability for a p-wave gap function 
$\Phi({\bf p},i\omega_n)$ transforming as $\sin(p_xa)$ 
(or $\sin(p_ya)$, the two being degenerate for 
a square lattice).

Figs.(2a),(3a) and (4a) show $T_c$ versus the dimensionless
coupling parameter $g^2\chi_0/t$ for several values of the square of the
inverse correlation length parameter $\kappa^2$ while figs.(2b),
(3b) and (4b) show $T_c$ versus $\kappa^2$ for several values
of the coupling parameter $g^2\chi_0/t$. The parameter
$\kappa^2$ can be varied experimentally by applying pressure 
to the samples. The $T_c$ versus $\kappa^2$ graphs can be 
interpreted as $T_c$ versus pressure plots, with the critical
pressure corresponding to the quantum critical point 
at $\kappa^2 = 0$. The critical temperature saturates, in 
the strong coupling limit, to a value of about $T_{sf}/30$ for 
values of $\kappa^2$ of 0.5 to 1.0. For long correlation lengths,
$T_c$ decreases. For fixed coupling constant $g^2\chi_0/t$, we find
that the Eliashberg renormalization factor 
$Z({\bf p},i\omega_n) = 1 - Im \Sigma({\bf p},i\omega_n)/\omega_n$ 
increases as $\kappa^2$ decreases and thus pair-breaking effects tend 
to cancel the stronger attraction as $\kappa\rightarrow 0$, 
leading to the reduction of the transition temperature. For short 
correlation lengths, $T_c$ is reduced as well since in that case 
the p-wave component of the pairing interaction becomes very small 
as it is nearly momentum independent for large 
values of $\kappa^2$. Figs. (2b),(3b) and (4b) show that for 
larger values of the characteristic spin-fluctuation frequency 
$T_{sf}$, the critical temperature is more sensitive to changes 
in $\kappa^2$. 

Our results for the mean-field transition temperature $T_c$ to 
a $d_{x^2-y^2}$ superconducting state ($\Phi({\bf p},i\omega_n)$ 
transforming as $\cos(p_xa)-\cos(p_ya)$) for antiferromagnetic 
spin-fluctuations are shown in fig.(5). Comparing with the 
results diplayed in fig.(3), one sees that for identical values 
of the characteristic spin-fluctuation temperature $T_{sf}$, the
d-wave transition temperature saturates to a value of about
$T_{sf}/2$ for values of $\kappa^2$ of 0.5 to 1.0, a factor
of ten or so larger than their p-wave counterparts. One also
observes from figs.(3a) and (5a) that $T_c$ saturates much more
rapidly to its largest value as $g^2$ is increased in the
antiferromagnetic case than it does for ferromagnetic 
spin-fluctuations. One sees from figs.(3b) and (5b) that the 
transition temperature is much less sensitive to changes in
$\kappa^2$ in the d-wave case than it is for p-wave 
superconductivity. As the inverse correlation length $\kappa^2$
is reduced, the mean-field $T_c$ is much more robust for
antiferromagnetic spin-fluctuations than for their ferromagnetic
counterparts, indicating that pair-breaking effects are not 
as damaging in the former case. The Eliashberg renormalization factor
$Z({\bf p},i\pi T)$ is shown in figs.(6) and (7) versus wavevector 
$\bf p$ for ferromagnetic and antiferromagnetic spin-fluctuations
for $\kappa^2 = 0.25$. The average of $Z({\bf p},i\pi T)$ over the
Fermi surface as a function of $\kappa^2$ is shown in fig.(8) for the
ferromagnetic case for several values of the coupling parameter
$g^2\chi_0/t$. We point out that even in the ferromagnetic case, $Z$ is 
strongly anisotropic around the Fermi surface when the coupling parameter
is small (fig.(6a)) and becomes more isotropic in the strong coupling
limit (fig.(7a)). The anisotropy for small coupling parameter can be 
understood as a density of states effect, since the smaller Fermi velocity
near the $(\pi/a,0)$ point can account for a larger value of $Z$ in this
region of the Brillouin Zone. These effects should matter less in the strong
couping limit. On the other hand, for antiferromagnetic spin-fluctuations, 
the anisotropy of $Z$ increases as the coupling parameter is increased 
(see figs.(6b) and (7b)). Finally, as shown in fig.(8) for nearly 
ferromagnetic systems, $Z$ increases rapidly and tends to diverge as the
inverse correlation length $\kappa \rightarrow 0$.

\section{Discussion}

The magnetic interaction potential, Eqs.~(\ref{chiML,Gap}) is attractive 
everywhere for the ferromagnetic case, but oscillates in space from 
attractive (odd sites) to repulsive (even sites) in nearly antiferromagnetic
metals. Since the average potential in the latter case tends to cancel, 
it may seem surprising at first sight that pairing is
so much more effective in nearly antiferromagnetic than ferromagnetic metals.
Part of the explanation lies in the fact that the inner product of the 
spins of two interacting quasiparticles, ${\bf s_1}\cdot {\bf s_2}$ 
that enters the pairing potential, is on average three times larger 
in magnitude for the spin singlet than the spin triplet 
state for spin $1\over 2$ particles (classically the 
expectation value would of course be the same in both cases). Thus the
ferromagnetic interaction potential, though everywhere attractive, is for 
this reason alone, three times weaker than the antiferromagnetic potential.
One can make this argument more quantitative and solve the 
Eliashberg equations for the nearly ferromagnetic metal assuming only 
the longitudinal spin-fluctuation mode contributes to the self-energy, 
setting the coupling parameter $g^2\rightarrow g^2/3$ in 
Eq.~(\ref{Sigma}) (the 'Ising' case). The results of the calculations 
for a spin-fluctuation temperature $T_{sf}$ equal to two thirds of the 
nearest neighbor hopping energy $t$ are shown in fig.(9) and to be 
compared with the results shown in figs.(3) and (5). 
While the critical temperatures of the nearly ferromagnetic 'Ising' 
metal are much higher than those of the the nearly ferromagnetic 
one for similar conditions, they do not quite match those of the 
nearly antiferromagnetic case. Therefore, the factor of three in 
the pairing potential is not the whole story. The extra factor of $q$ 
from Landau damping in Eq.~(\ref{ferro}) leads to greater incoherent 
scattering for a nearly ferromagnetic than antiferromagnetic metal, 
and hence to a reduced $T_c$. We have also solved the Eliashberg 
equations for the nearly ferromagnetic 'Ising' metal without Landau 
damping (with $\eta(\widehat{q}) = T_{sf}\widehat{q}_{+}$, 
in Eq.~(\ref{ferro})). The results for the same value of $T_{sf}$ are shown in 
fig.(10). One might have expected that the 'Ising' case without the 
Landau damping would lead to transition temperatures for the purely
attractive potential and p-wave pairing that are much higher than
from the spatially oscillatory potential and d-wave pairing. That
this is not the case, as may be seen by comparing figs.(5) and (10),
can be understood when one takes into account of the effects of 
retardation that restricts scattering to states within a narrow 
range of wavevectors near the Fermi surface. This implies that the pair 
wavefunction tends to oscillate in space with wavevector of the order of 
$k_F$ and the probability distribution with wavevector $2k_F$, i.e 
with a wavevector comparable to that of the magnetic interaction potential 
itself. Furthermore, the maxima of the probability appear near the minima 
of the potential along the square axes, while in the d-wave state, the 
probability vanishes alltogether along the diagonals where the interaction 
is everywhere repulsive. In this case the effect of the repulsive regions 
is small and the gain achieved with a purely attractive potential with 
otherwise similar properties is not as great as might have naively 
been suspected.

Beside this there remains at least one more significant difference
between the ferromagnetic and antiferromagnetic cases that may be
relevant to pair formation but is not readily quantified. In the
latter case the mass renormalization is much more anisotropic than in the
former and is strongest at points on the Fermi surface (the 'hot spots')
connected by the antiferromagnetic wavevector. This anisotropy may lead 
to strong coupling effects which on the whole are less damaging to
pairing than in the corresponding ferromagnetic case where essentially
all the points of the Fermi surface are equivalent.

Taken together, these effects confer a very considerable advantage 
for pairing to the nearly antiferromagnetic versus ferromagnetic metals
that have otherwise comparable properties. Further considerations also
lead to an advantage of quasi 2D over 3D metals. The average of the 
spin-fluctuation frequency in the Brillouin Zone tends to be larger 
in 2D than in 3D. This favors enhanced incoherent scattering, and 
hence reduced $T_c$. However, it also leads to an enhanced pairing 
energy and greater robustness against impurities and the effects of 
competing channels of interactions. We expect that these latter 
considerations will normally tend to dominate and hence favor 
quasi 2D over 3D systems, under otherwise similar conditions, 
and provided that corrections to the mean-field solutions are not 
important.

Within the magnetic interaction model in the mean-field approximation, thus, 
the highest $T_c$ is expected to arise in quasi 2D metals with high $T_{sf}$
and on the border of a continuous antiferromagnetic transition (when 
the magnetic correlation wavevector $\kappa \rightarrow 0$ as $T \rightarrow 0$). 
Interestingly these conditions are well satisfied in the copper oxides but 
much less so in the heavy fermion and organic compounds (see e.g. refs
~\cite{Jerome} and ~\cite{Mathur} respectively). In the heavy 
fermions $T_{sf}$ happens to be low because the f electrons produce narrow 
bands, while in the organics $T_{sf}$ is small because the carrier 
concentration is low. Thus one expects, and indeed one finds, much lower
$T_c$'s in these materials than in the cuprates.

The calculations also predict that magnetically mediated superconductivity 
should be a general phenomenon occuring on the boundary of a continuous
magnetic transition, in both ferromagnets and antiferromagnets and in 
quasi 2D and 3D compounds. This may not be observed in practice, however, 
due to pair breaking effects of impurities and other interaction 
channels not considered here explicitly. In cases when the magnetic 
transistion is not abrupt and $\kappa$ can be made 
arbitrarily small at low temperatures, the magnetic 
interaction potential may overwhelm these other effects and, at least 
in the nearly antiferromagnetic case where the mean-field $T_c$ 
appears to remain finite as $\kappa \rightarrow 0$, superconductivity 
may survive in a narrow range of lattice densities near the critical density
where the magnetic order is continuously quenched. 

The magnetic interaction model and the mean-field approximation for $T_c$ 
might be expected to apply most successfully in nearly magnetic metals where
$T_c$ is small compared to the electronic bandwidth and $T_{sf}$. 
The nearly ferromagnetic quasi 2D metal $Sr_2RuO_4$ that orders in a 
spin-triplet p-wave state only at very low temperatures (below $1.5^\circ K$),
\cite{Agterberg,Rice,Baskaran,Maeno,Luke,Rieseman,Ishida,Mackenzie,Yoshida} 
would therefore seem an ideal candidate for comparison between 
theory and experiment. The calculations presented in this paper provide 
a first step toward such a comparison. The next will be to build a realistic
model of $\chi({\bf q},\omega)$ from NMR and neutron scattering measurements,
or from the numerical calculations now in progress\cite{Mazin}. 
Preliminary evidence suggests that $Sr_2RuO_4$ may be close both to 
ferromagnetism and antiferromagnetism\cite{Mazin2}. The competition between 
these two tendencies, along with the comparatively small magnitude of 
$<{\bf s_1}\cdot {\bf s_2}>$ in the observed spin-triplet state and other 
features as discussed above, may help to account for the much lower $T_c$ 
in this layer perovskite oxide compared with that of the cuprates.

We note that our calculations may be expected to break down when the
mass renormalization becomes large at high values of the coupling constant
or at small $\kappa$ near the critical point for magnetic order. Also it 
should fail when the superconducting coherence length becomes small compared
with the average spacing between Cooper pairs, i.e. for sufficiently 
high $T_c$ or in {\it strictly} 2D where there is no true long-range order
at finite temperature. The latter condition is not readily reached in many
of the known quasi 2D systems.

Finally, we emphasize that our model for the magnetic interaction does not
include any possible spin-gap formation. For this reason alone, it is not
expected to apply near the metal-insulator phase boundary in the 
cuprates\cite{PWA}.

\section{Conclusions}

We have contrasted the predictions for the superconducting 
transition temperature for magnetically mediated superconductivity for
nearly ferromagnetic versus nearly antiferromagnetic metals in quasi 2D.
The calculations are based on a single Fermi surface sheet, and a 
conventional form for the magnetic interaction arising from the exchange 
of spin fluctuations treated in the mean-field Eliashberg theory. The
dominant $\bf q$  and $\omega$ dependence of this interaction is assumed
to arise from the dynamical wavevector dependence of the susceptibility, 
and thus the interaction vertex is taken to be a phenomenological constant.
In principle the latter quantities may be inferred independently from inelastic
neutron scattering and for example the temperature dependence of the 
resistivity in the normal state. The mean-field Eliashberg theory is 
expected to break down when, for example, $T_c$ is so high that the 
superconducting coherence length becomes small and less than the typical 
spatial separation of Cooper pairs. It may also fail in the immediate 
vicinity of the critical density when magnetic order is quenched 
continuously and the quasiparticle density of states tends to become 
singular. Here the electron quasiparticle framework underpinning the 
mean-field Eliashberg model may break down in an essential
and non-trivial fashion.

Within the range of validity of our calculations we may conclude that, for the 
same set of dimensionless parameters, the p-wave triplet pairing in nearly 
ferromagnetic metals is much less robust than the d-wave singlet pairing in 
the corresponding nearly antiferromagnetic metals. For values of $T_{sf}$ 
that are typical of d metals in the layered perovskites, we predict a maximum
of $T_c$ versus $\kappa$ of the order of $100^\circ K$ in the latter but
typically one or more orders of magnitude less than this in the former.
The reasons for the dramatic difference are discussed in section IV.

The pair breaking effects of impurities and of competing interaction channels
can lead to substantially lower values than the above in real materials. These
effects may, however, be mitigated by reducing $\kappa$ via some external
control parameter such as pressure and hence enhancing 
the magnetic pairing energy.

The calculations are intriguing in the light of the d-wave singlet state 
observed in the cuprates with strongly enhanced antiferromagnetic 
susceptibilities and $T_c$'s of the order of $100^\circ K$, versus the
p-wave triplet state found in the ruthanate $Sr_2RuO_4$ that is close to
ferromagnetic order and has a much lower $T_c$ (of the order of $1^\circ K$).

The maximum of $T_c$ versus $\kappa$ in $Sr_2RuO_4$ in the triplet state 
is not yet known, and may well be higher than that measured at ambient 
pressure. A more complete description of $Sr_2RuO_4$ must await realistic 
modelling of the dynamical susceptibility which may reflect not only 
ferromagnetic but also competing antiferromagnetic tendencies. The latter 
may be subdominant at ambient pressure but may be highly sensitive to 
lattice spacing. It would also be interesting to investigate more closely the
effect of the additional Fermi surface sheets in $Sr_2RuO_4$.
An experimental study of the variation of $T_c$ vs 
$\kappa$ in this system, that satisfies to a greater extent than cuprates 
the condition $T_c \ll T_{sf}$, would provide a vital 
test of the theory of magnetically mediated superconductivity. 
Such a study would be feasible in $Sr_2RuO_4$ if it orders 
ferromagnetically at positive pressure\cite{Yoshida}, and in the 
isostructural and isoelectronic compounds $Ca_2RuO_4$ and $Sr_2FeO_4$ 
that are expected to become similar to $Sr_2RuO_4$ at very high pressure.

Finally we reiterate that our calculations suggests that one should 
look for elevated transition temperatures in systems in 
which (i) $T_{sf}$ is high, i.e. the electron density is 
not too low and effective band mass not too high, 
(ii) the lattice or carrier density can be tuned to the vicinity of a 
magnetic critical point in the metallic state, (iii) the electronic
structure is quasi 2D rather than 3D, and (iv) antiferromagnetism
(or 'Ising' ferromagnetism) is favored over ferromagnetism. A
considerable number of candidate materials for further study of the 
predictions of the magnetic pairing model would seem to be available given
current material fabrication and high pressure technology. The experimental
investigation of such systems, whether or not they prove to yield high
transition temperatures, should help us to improve our understanding of
magnetic pairing and perhaps also shed light on the more exotic models\cite{PWA}
for normal and superconducting states that have been proposed for 
highly correlated electronic systems.

\section{Acknowledgments}

We would like to thank P. Coleman, S.R. Julian, P.B. Littlewood, A.J. Millis, 
A.P. Mackenzie, D. Pines, D.J. Scalapino and M. Sigrist for discussions on 
this and related topics. We acknowledge the support of the EPSRC and the 
Royal Society.

\begin{figure}
\caption{The Fermi surface for t' = 0.45t and electron density n = 1.1.
\label{fig1}}
\end{figure}

\begin{figure}
\caption{The mean-field critical temperature $T_c$ to 
the p-wave superconducting state versus $g^2\chi_0/t$ for 
$\kappa^2 = $ 0.25, 0.50, 1.0, 2.0, 3.0, 4.0 (a)
and versus $\kappa^2$ for $g^2\chi_0/t = $ 60, 30, 20, 10, 5 (b).
The characteristic spin-fluctuation temperature is $T_{sf} = 0.33t$ with
$\kappa_0^2 = 24$.
\label{fig2}}
\end{figure}

\begin{figure}
\caption{The mean-field critical temperature $T_c$ to 
the p-wave superconducting state versus $g^2\chi_0/t$ for 
$\kappa^2 = $ 0.25, 0.50, 1.0, 2.0, 3.0, 4.0 (a)
and versus $\kappa^2$ for $g^2\chi_0/t = $ 60, 30, 20, 10, 5 (b).
The characteristic spin-fluctuation temperature is $T_{sf} = 0.67t$ with
$\kappa_0^2 = 12$.
\label{fig3}}
\end{figure}

\begin{figure}
\caption{The mean-field critical temperature $T_c$ to 
the p-wave superconducting state versus $g^2\chi_0/t$ for 
$\kappa^2 = $ 0.25, 0.50, 1.0, 2.0, 3.0, 4.0 (a)
and versus $\kappa^2$ for $g^2\chi_0/t = $ 60, 30, 20, 10, 5 (b).
The characteristic spin-fluctuation temperature is $T_{sf} = 1.33t$ with
$\kappa_0^2 = 6$.
\label{fig4}}
\end{figure}

\begin{figure}
\caption{The mean-field critical temperature $T_c$ to 
the d-wave superconducting state versus $g^2\chi_0/t$ for 
$\kappa^2 = $ 0.25, 0.50, 1.0, 2.0, 3.0, 4.0 (a)
and versus $\kappa^2$ for $g^2\chi_0/t = $ 60, 30, 20, 10, 5 (b).
The characteristic spin-fluctuation temperature is $T_{sf} = 0.67t$ with
$\kappa_0^2 = 12$.
\label{fig5}}
\end{figure}

\begin{figure}
\caption{The Eliashberg renormalization factor $Z({\bf p},i\pi T) = 1 -
Im\Sigma({\bf p},i\pi T)/\pi T$ versus wavevector ${\bf p}$ for 
ferromagnetic (a) and antiferromagnetic spin-fluctuations (b) 
for $g^2\chi_0/t = 5$, $\kappa^2 = 0.25$ and $T = 0.00625t$. 
The characteristic spin-fluctuation temperature is $T_{sf} = 0.67t$ 
and $\kappa_0^2 = 12$.
\label{fig6}}
\end{figure}

\begin{figure}
\caption{The Eliashberg renormalization factor $Z({\bf p},i\pi T) = 1 -
Im\Sigma({\bf p},i\pi T)/\pi T$ versus wavevector ${\bf p}$ for 
ferromagnetic (a) and antiferromagnetic spin-fluctuations (b) 
for $g^2\chi_0/t = 30$, $\kappa^2 = 0.25$ and $T = 0.00625t$. 
The characteristic spin-fluctuation temperature is $T_{sf} = 0.67t$ 
and $\kappa_0^2 = 12$.
\label{fig7}}
\end{figure}

\begin{figure}
\caption{The Eliashberg renormalization factor $Z({\bf p},i\pi T) = 1 -
Im\Sigma({\bf p},i\pi T)/\pi T$ averaged over the Fermi surface for 
ferromagnetic spin-fluctuations versus $\kappa^2 = 0.25$ for 
$g^2\chi_0/t =$ 5, 10 and 30, and $T = 0.003125t$. 
The characteristic spin-fluctuation temperature is $T_{sf} = 0.67t$ 
and $\kappa_0^2 = 12$.
\label{fig8}}
\end{figure}

\begin{figure}
\caption{The mean-field critical temperature $T_c$ to 
the p-wave 'Ising' superconducting state versus $g^2\chi_0/t$ for 
$\kappa^2 = $ 0.25, 0.50, 1.0, 2.0, 3.0, 4.0 (a)
and versus $\kappa^2$ for $g^2\chi_0/t = $ 60, 30, 20, 10, 5 (b).
The characteristic spin-fluctuation temperature is $T_{sf} = 0.67t$ with
$\kappa_0^2 = 12$. For the smae value of the coupling parameter the
effective mass for 'Ising' p-wave pairing is much lower than for the
standard p-wave state. That explains the more rapid drop of $T_c$
as $\kappa \rightarrow 0$ in fig.(3) than in the above figure.
\label{fig9}}
\end{figure}

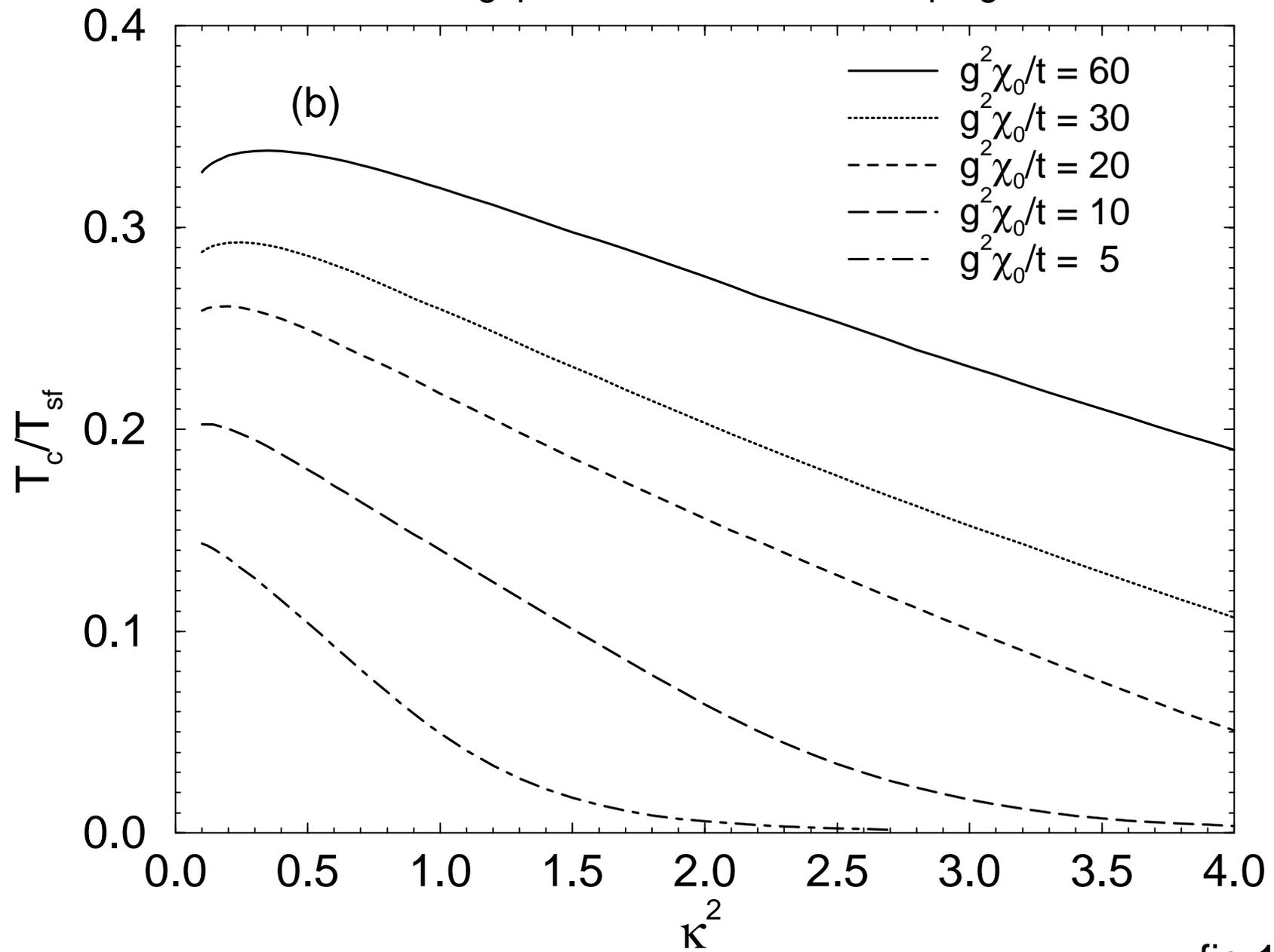
\begin{figure}
\caption{The mean-field critical temperature $T_c$ to 
the p-wave 'Ising' superconducting state without Landau damping
versus $g^2\chi_0/t$ for $\kappa^2 = $ 0.25, 0.50, 1.0, 2.0, 3.0, 4.0 (a)
and versus $\kappa^2$ for $g^2\chi_0/t = $ 60, 30, 20, 10, 5 (b).
The characteristic spin-fluctuation temperature is $T_{sf} = 0.67t$ with
$\kappa_0^2 = 12$.
\label{fig10}}
\end{figure}

\end{document}